\documentclass{elsart}
\def\lkh{{L}}
\def\ss{{S}}
\def\bb{{B}}
\input psfig
\long\def\simplex#1#2#3#4{
\begin{figure}[#1]
   \begin{center}
   \quad\\[-2.5cm]
   \quad\vskip  2.0 cm
   \hbox{
   \quad 
   \parbox[t]{6.5cm}{ \psfig{figure=#2,width=8.5cm}
   \caption[]{\small  \label{fig:#3} #4 } }
   }
   \quad
   \end{center} 
\end{figure}
}
\long\def\sixplex#1#2#3#4#5#6#7#8#9{
\begin{figure}[#1]
   \begin{center}
   \quad\\[-2.5cm]
   \quad
   \hbox{\hskip -2cm
   \quad 
   \parbox[t]{6.5cm}{ \psfig{figure=#2,width=8.5cm}
   } 
   \quad
   \parbox[t]{6.5cm}{ 
   \psfig{figure=#3,width=8.5cm} }
   }
   \quad
   \end{center}
   \begin{center}
   \quad\\[-2.5cm]
   \quad
   \hbox{\hskip -2cm
   \quad 
   \parbox[t]{6.5cm}{ \psfig{figure=#4,width=8.5cm}
   } 
   \quad
   \parbox[t]{6.5cm}{ 
   \psfig{figure=#5,width=8.5cm} }
   }
   \quad
   \end{center}
   \begin{center}
   \quad\\[-2.5cm]
   \quad
   \hbox{\hskip -2cm
   \quad 
   \parbox[t]{6.5cm}{ \psfig{figure=#6,width=8.5cm}
   } 
   \quad
   \parbox[t]{6.5cm}{ 
   \psfig{figure=#7,width=8.5cm} }
   }
   \caption[c]{\small \label{fig:#8} #9 }
   \quad
   \end{center}
\end{figure}
}
\begin{document}
\begin{frontmatter}
\title{Estimation of Upper Limits Using a Poisson Statistic}
\author{Ilya Narsky\thanksref{addr}}
\thanks[addr]{Tel.: 1 214 768 1299; Fax: 1 214 768 4095; E-mail:
narsky@mail.physics.smu.edu.} 
\address{Physics Department, Southern Methodist University, Dallas,
TX, 75275-0175, USA}
%\thanks{Supported by the Department of Energy}

\begin{abstract}
Bayesian, classical, and extended maximum likelihood approaches to
estimation of upper limits in experiments with small numbers of signal
events are surveyed. The discussion covers only experiments whose
outcomes are well described by a Poisson statistic.
A new approach, based on the statistical
significance of a signal 
rather than on the number of events in the signal region, is
proposed. A toy
model and an example of a recent search for the lepton number
violating decay $\tau\to\mu\gamma$ are used to
illustrate application of the discussed techniques.
\end{abstract}
\end{frontmatter}

%\begin{spacing}{2.0}
\section{Introduction}

Searches for rare signals often fail to detect a signal of
sufficient statistical significance and thus
face the need to set an upper limit on the signal rate. Unfortunately,
there is no standard prescription for setting such limits, and a
number of techniques have been employed in the past to meet
this challenge. 
This problem is particularly important for the particle physics
community. Upper limits on rare and forbidden decays can provide
valuable constraints on physics beyond the Standard Model of
Electroweak Interactions.

This paper is stimulated by the observation that
often authors do not pay enough attention to the choice 
of a procedure for upper limit estimation.
Furthermore, many experimentalists do not
provide a sufficient description of the procedure used in
their analysis technique, assuming that the reader is smart enough to figure
out the details. An example given in Section~\ref{sec:class} shows that, 
in some situations, the choice
of a procedure can change the value of an upper limit {\em by
an order of magnitude}, which suggests that this problem should not be
taken lightly.

In Section~\ref{sec:survey} the most popular approaches to
upper limit estimation are surveyed and criticism is provided, where
applicable. In 
Section~\ref{sec:ss} definitions of the statistical
significance of a signal are discussed and a new technique, which estimates
upper limits using these definitions, is proposed. Unlike
commonly used procedures, this approach does not just count the number of
events in the signal region, but takes into account statistical
fluctuations of the background as well. In Section~\ref{sec:toy_model}
various techniques are compared using a toy model. In 
Section~\ref{sec:cleo} an example of a recent
search for the lepton number violating decay
$\tau\to\mu\gamma$~\cite{mugamma}, performed by the CLEO
collaboration, is given. 

The emphasis is made on estimation of upper limits, though the
discussion can be easily generalized to include construction of
confidence intervals whose lower bound is not constrained to zero.

There is no discussion of systematic effects in this paper. Throughout
the paper the number of observed events, the expected
background rate and, if applicable, the coordinates of
observed events are assumed to be measured with perfect accuracy.

\section{Commonly Used Techniques for Upper Limit Estimation}
\label{sec:survey}

\subsection{Bayesian Approach}
\label{sec:bayes}

In a Bayesian approach one has to assume a {\em prior} probability
density function (pdf) of an unknown parameter and then perform an
experiment to update the 
prior distribution. A prior pdf reflects knowledge that is available to an
experimentalist {\em before} an experiment is performed. The updated
prior is called the {\em posterior} pdf and is used to draw inference
on the unknown parameter. 
This updating is done with
the use of Bayes' Rule. For the moment let us ignore the issue of
background, i.e., let us assume that the background rate is measured
very accurately and thus can be treated as a known constant. Then the
only unknown parameter is the signal rate $s$. Bayes' Rule gives:
\begin{equation}
\label{eq:poster}
\pi(s|n) = \frac{ f(n|s) \pi(s) }{ \int_{0}^{\infty} f(n|s) \pi(s) ds
}\ ,
\end{equation}
where $n$ represents the number of observed events,  
$f(n|s)$ is the conditional probability to observe $n$ events, given
the signal rate $s$, $\pi(s)$ is the prior pdf, and $\pi(s|n)$ is the
conditional posterior pdf. 
%The obtained posterior pdf gives full
%information about the parameter $s$. 
Now, for any given confidence level, one can compute a 
Bayesian confidence interval for the signal rate $s$. For upper limit
estimation the natural choice 
of the Bayesian confidence interval is of the form $(0,s_0)$. Here, $s_0$
denotes an upper 
limit and can be found from the equation:
\begin{equation}
\label{eq:bayes_cl}
1-\alpha = \int_0^{s_0} \pi(s|n) ds\ .
\end{equation}
The confidence
level is denoted by $(1-\alpha)$ following the conventional
statistical notation. A nice feature of the Bayesian 
approach is that the zero value of an upper limit $s_0$ always corresponds
to the zero value for the confidence level $(1-\alpha)$. As you will
see below, this is not necessarily true for the classical approach.

The most important step here is to define a prior
distribution of the parameter. Naturally, this is the step which
brings most of controversy (and sometimes confusion) into Bayesian
methods. Many statistical textbooks treat a prior pdf as a purely
subjective assumption which is based on experimenter's belief. At the
same time there are authors~\cite{jeffr1,jeffr2,jaynes,prosp1} who
advocate the objectivity of prior assumptions. In particular,
Jaynes~\cite{jaynes} stated the ``basic desideratum'' of the objective
approach as follows: {\em
in two problems where we have the same prior information, we should
assign the same prior probabilities}. These authors have offered a
number of mathematical procedures that can be used to convert
prior knowledge into an exact formula for the prior distribution.
Another important question is
whether one should assume an {\em informative} prior, i.e., a prior
which incorporates results of previous experiments, or a {\em
non-informative} prior, i.e., a prior which claims total
ignorance. Naively, it would seem unnatural to use non-informative
priors since the power of the Bayesian approach comes from the fact
that we update our prior knowledge rather than start every analysis
from scratch. In particle physics the major
objection against informative priors is based on the following
argument: if we assume a prior which incorporates results of earlier
experiments, then our experiment will not be independent of those and
thus we will not be able to combine our
results with the results of previous experiments just by taking a
weighted average. So far, particle physicists have been
largely ignoring informative priors. Thus, the discussion below covers
only those Bayesian methods that assume a non-informative prior pdf
for the positive parameter of a Poisson distribution.

In the absence of background the conditional pdf $f(n|s)$ is given by: 
\begin{equation}
\label{eq:poiss}
f(n|s) = e^{-s} \frac{s^n}{n!}\ .
\end{equation}
Bayes and Laplace~\cite{bayes,laplace}, who pioneered statistical
research employing 
Bayesian methods, stated that the non-informative prior for any
parameter must be flat. This conclusion was not based on any strict
mathematical argument, it was merely a product of their
intuition. Modern advocates~\cite{hel1,hel2,hel3,hel4} of this
approach do not offer any mathematical explanation either, they just
consider a flat prior pdf as the most natural choice one can make. As
natural as this assumption may seem, there are several objections. The
most obvious argument is that if one can assume a flat distribution of
an unknown parameter, then one can also assume a flat distribution
for any function of this parameter, and these two assumptions are
clearly not identical. 
If an experiment is measuring the mean of the Poisson
statistic~(\ref{eq:poiss}), then one can argue that 
%If the mean of the Poisson statistic~(\ref{eq:poiss}) is being
%measured, then one can argue that 
%In case of a Poisson statistic this objection
%can be waived, since 
the most natural candidate for the unknown
parameter is the signal rate $s$, and we have no physical reasons to consider
any functions of this parameter except the parameter itself. In other
experiments the situation may be not so simple. For example, if an
experiment is measuring a neutrino mass, then typically the measured
quantity is not the neutrino mass itself, but the neutrino mass
squared. In this situation it's not clear whether one should choose mass or
mass squared as a candidate for the flat
distribution. Jeffreys~\cite{jeffr1} resolved this problem by
introducing an {\em invariate} prior pdf $1/\theta$ which, he
stated, was a valid non-informative prior for all problems where the
unknown parameter $\theta$ could vary from $0$ to $+\infty$. His choice was
mostly motivated by the fact that $d\theta/\theta \propto
d\theta^n/\theta^n$, i.e., the pdf of $\theta$ stays invariant under
any power transformation. All non-power functions of $\theta$ were
rejected by Jeffreys as non-physical under the assumption that the
parameter $\theta$ is a dimensional quantity. This argument was put on
a more rigorous mathematical basis by Jaynes~\cite{jaynes} who
stipulated that a prior pdf has to stay invariant under any symmetry
transformation\footnote{Strictly speaking, Jaynes' argument is not
applicable to the 
Poisson distribution~(\ref{eq:poiss}) with a non-dimensional mean rate
$s$. Originally Jaynes' non-informative prior was derived
for the Poisson pdf $f(n|s) = e^{-st}(st)^n/n!$, where $n$ is the number of
counts, $t$ is the counting time, and $s$ is the signal rate per unit time.
However, the requirement of dimensionality seems to be somewhat
arbitrary. For example, in the above formula 
$t$ can be the amount of statistics
accumulated in the experiment and $s$ can be the signal rate normalized to
this amount of statistics. A prior pdf has to stay invariant under the
transformation $s'=qs$, $t'=t/q$, i.e., Jaynes' argument is applied.}
that does not change the physics of an experiment. His
conclusion was similar to that of Jeffreys, namely that the
non-informative prior for a Poisson statistic~(\ref{eq:poiss}) has to
be proportional to $1/s$. 
An alternative approach was developed by Box
and Tiao~\cite{box} who introduced the notion of a {\em
data-translated} likelihood. In their approach a prior pdf is
non-informative if the location, but not the shape, of the
corresponding posterior likelihood\footnote{As usually, a likelihood function 
is defined by swapping argument and parameter in the expression of the
corresponding pdf.} is determined by the
unknown parameter. Thus, the location of the posterior likelihood is
completely defined by data, while its shape had been determined before
the data were seen (hence the name ``data-translated'' likelihood). It
may not be possible to construct a data-translated likelihood for
every distribution. In this case one can use Taylor expansion to
construct an approximate data-translated likelihood. In particular,
a non-informative prior which produces an approximate data-translated
likelihood for the Poisson statistic~(\ref{eq:poiss}) is given by
$1/\sqrt{s}$. 

In the presence of background, the Poisson pdf~(\ref{eq:poiss}) has to be
modified to account for the non-zero background rate $b$:
\begin{equation}
\label{eq:poiss_b}
f(n|s) = e^{-(s+b)} \frac{(s+b)^n}{n!}\ .
\end{equation}
If the background rate is accurately measured, it can be treated
as a known constant. In this case an argument similar to that of
Ref.~\cite{jaynes} gives $1/(s+b)$ and
an argument similar to that of Ref.~\cite{box} gives $1/\sqrt{s+b}$ for the
non-informative prior. In general, for the prior pdf 
\begin{equation}
\label{eq:prior}
\pi(s) \propto \frac{1}{(s+b)^m}\ ;\ \ \ \ 0\leq m\leq 1\ ;
\end{equation}
the posterior distribution is given by
\begin{equation}
\label{eq:poster_1sm}
\pi(s|n) = \frac{ e^{-(s+b)} (s+b)^{n-m} }{ \Gamma(n-m+1,b) }\ ,
\end{equation}
where
\begin{equation}
\Gamma(p,\mu) = \int_{\mu}^{\infty} s^{p-1} e^{-s} ds ;
\ \ p>0 ;\ \ \mu>0 ;
\end{equation}
is an incomplete gamma-function.
Substituting the posterior pdf~(\ref{eq:poster_1sm}) into
Eqn.~(\ref{eq:bayes_cl}), we obtain:
\begin{equation}
\label{eq:bayes_ul}
1-\alpha = 1 - \frac{ \Gamma(n-m+1,s_0+b) }{ \Gamma(n-m+1,b) }\ .
\end{equation}
%To be well-defined, a gamma-function needs to have its first argument
%strictly positive. 
A gamma-function is not well-defined if its first argument is equal to
zero or a negative integer.
Thus, at $n=0$ (no events observed) and $m=1$ (the
$1/(s+b)$ prior) Eqn.~(\ref{eq:bayes_ul}) fails to find an upper limit.
Under the flat prior $m=0$,
Eqn.~(\ref{eq:bayes_ul}) turns into the formula
\begin{equation}
\label{eq:pdg96}
1-\alpha = 1 -
\frac{ \sum_{k=0}^{n} e^{-(s_0+b)} \frac{(s_0+b)^k}{k!} }
{ \sum_{k=0}^{n} e^{-b} \frac{b^k}{k!} }\ ,
\end{equation}
which was adopted by Particle Data Group~\cite{pdg96}. Here and below,
$0^0$ is by definition equal to one. 

If the background rate is measured with large uncertainty, then the
situation is more complicated. The usual approach is to modify
the Poisson statistic~(\ref{eq:poiss_b}) as:
\begin{equation}
f(n|s) = \int_0^{+\infty} e^{-(s+b)} \frac{(s+b)^n}{n!} f(b) db\ ,
\end{equation}
where $f(b)$ is the measured or predicted pdf of the background rate
$b$. The argument of the previous paragraph, leading to the $1/(s+b)$
and $1/\sqrt{s+b}$ expressions for the non-informative prior, is not
necessarily valid in this case. A more consistent Bayesian approach
would be to derive a joint non-informative prior $\pi(s,b)$, thus
treating both the signal and background rates as unknown
parameters. An example of such derivation is given in
Ref.~\cite{prosp1}. A discussion of intricacies, that might arise
from inclusion of background uncertainty into the upper
limit calculation, is beyond the material covered in this paper.

\subsection{Classical Approach}
\label{sec:class}

The classical (or frequentist) approach is traditionally interpreted
in the following way: if a $(1-\alpha)$ classical confidence set is constructed
for the unknown parameter, then the probability for this confidence
set to cover the true value of the parameter equals $(1-\alpha)$, that
is, if an infinite number of identical independent
experiments is performed and for each of these a
$(1-\alpha)$ confidence 
set is constructed, then $100(1-\alpha)\%$ of these confidence sets will and
$100\alpha\%$ of these confidence sets will not contain the true
value of the parameter. Rules for construction of classical intervals
were outlined in a famous work~\cite{neyman} by Neyman. Here I treat
the terms ``classical'' and ``frequentist'' as 
equivalent. The implication is that the classical approach is
inevitably connected to the concept of an identical independent
experiment. Sometimes this concept is misunderstood, and are curious
attempts to treat certain problems in the ``classical'' 
vein while, in fact, this treatment has nothing to do with the
frequentist approach. An example of such misunderstanding is shown in
this Section. 

A confidence set for the unknown signal rate $s$ is typically
constructed~\cite{garwood} as a confidence interval $s_1\leq s\leq s_2$
satisfying:
\begin{equation}
\label{eq:neyman}
1 - \alpha = \sum_{k=n}^{\infty} f(k|s_1) + \sum_{k=0}^{n} f(k|s_2)\ .
\end{equation}
For upper limit estimation it is natural to consider intervals $0\leq
s\leq s_0$, where $s_0$ is the value of an upper
limit. Eqn.~(\ref{eq:neyman}) is thus reduced to:
\begin{equation}
\label{eq:class_ul}
1 - \alpha = 1 - \sum_{k=0}^{n} e^{-(s_0+b)} \frac{(s_0+b)^k}{k!}\ ,
\end{equation}
where $f(k|s_0)$ was replaced by its definition~(\ref{eq:poiss_b}).

Eqn.~(\ref{eq:class_ul}) looks similar to the Bayesian
formula~(\ref{eq:pdg96}), except that the denominator is now absent.
The denominator in Eqn.~(\ref{eq:pdg96}) represents the probability of
observing $n$ or less background events in the signal
region, and thus, it is always less than 1 except at
$b=0$. Therefore, for the non-zero 
background rate, i.e., $b\neq 0$, the classical 
approach~(\ref{eq:class_ul}) always provides a smaller value of an
upper limit than the Bayesian approach~(\ref{eq:pdg96}) with a flat prior.
The difference becomes significant when the denominator in
Eqn.~(\ref{eq:pdg96}) is small, i.e., when the number $n$ of events
observed in the signal region is small as compared to the expected
background rate $b$. For example, for the observed number of
events in the signal region $n=3$ and the expected background rate
$b=6.5$ the classical 
approach~(\ref{eq:class_ul}) gives an upper limit $s_0=0.18$ at 90\% confidence
level, while the Bayesian approach~(\ref{eq:pdg96}) with a flat
prior gives $s_0=3.39$, i.e., a difference of more than an order of
magnitude. Another
feature of Eqn.~(\ref{eq:class_ul}) is that now a zero upper limit:
$s_0=0$, does not give you a zero confidence level. When the expected
background rate $b$ is large, one can achieve a situation when
$1-\alpha < 1-\sum_{k=0}^{n} e^{-b} b^k/k!$, i.e.,
formula~(\ref{eq:class_ul}) gives an 
unphysical negative value of
an upper limit. Advocates of the Bayesian method consider this
as an undesirable feature of the classical approach. But the failure
to set an upper limit is not necessarily a bad feature. It just
implies that the observed
outcome of an experiment is highly improbable, and one should
question the experimental technique that was used to measure the 
signal and the background.

Zech~\cite{zech} made an attempt to derive the Bayesian
formula~(\ref{eq:pdg96}) 
using the classical approach. To arrive at Eqn.~(\ref{eq:pdg96}),
Ref.~\cite{zech} postulates that the number of background events cannot
exceed the number $n$ of observed events and therefore one has to
renormalize the probability~(\ref{eq:class_ul}) according to this constraint.
This approach misunderstands the concept of frequentist
coverage. Eqn.~(\ref{eq:class_ul}) answers the question: what is the
probability of observing $n$ or more events in an {\em
identical independent} experiment? This implies that the number $n$ of
events cannot be fixed at any particular value. If you fix $n$, then
you confine yourself to this particular 
drawing; the concept of an identical independent experiment is
therefore inapplicable. In this case
a binomial distribution should be used instead of a Poisson
pdf. 
%See Section~\ref{sec:b_vs_p} for related discussion.

Confidence intervals are not uniquely defined by Eqn.~(\ref{eq:neyman})
unless one imposes specific criteria on their construction. 
In the situation discussed above, the uniqueness was
introduced by requiring that the confidence set must be an interval
$(0,s_0)$. One can choose 
another requirement and obtain a different
confidence set. In general, if the problem is reduced to one variable
and one unknown parameter, then confidence intervals are obtained via
construction of confidence belts~\cite{eadie}. The idea behind this
technique is the following. For every assumed value of the signal rate
$s$, one finds an acceptance interval $n_1(s)\leq n\leq n_2(s)$ which
satisfies:
\begin{equation}
\label{eq:c_belt}
1 - \alpha = \sum_{n=n_1}^{n_2} f(n|s)\ .
\end{equation}
Due to the discrete nature of the Poisson distribution, it is usually
impossible to find values of $n_1$ and $n_2$ that satisfy
Eqn.~(\ref{eq:c_belt}). The standard solution is to stay on the
conservative side and to search for an acceptance interval that gives
at least the required coverage:
\begin{equation}
\label{eq:c_belt_2}
1 - \alpha \leq \sum_{n=n_1}^{n_2} f(n|s)\ .
\end{equation}
The obtained functions $n_1(s)$ and $n_2(s)$ define two curves on the
$s$-vs-$n$ plane. Then, for every value of $n$, one obtains an interval
$s_1(n)\leq s\leq s_2(n)$ which lies between these two curves. The
obtained interval is a $(1-\alpha)$ confidence interval for the specific
value of $n$. The order in which values of $n$ are added to the
acceptance region $(n_1,n_2)$ for a specific value of $s$ is called an
{\em ordering principle}. Thus, every ordering principle corresponds
to a specific set of confidence intervals $(s_1,s_2)$ indexed by the
observation variable $n$. The well-known examples are an old
paper~\cite{crow} by Crow and Gardner and a recent
paper~\cite{feldman} by Feldman and Cousins.
Ref.~\cite{crow}
minimizes the length of the acceptance interval $(n_1,n_2)$ defined by
Eqn.~(\ref{eq:c_belt_2}). Ref.~\cite{feldman} employs an ordering
principle based on likelihood ratios. The latter was adopted by the
last release of the Particle Data Group Review~\cite{pdg98}. 
In both approaches an experimentalist does not have to decide
whether she/he wants to quote an upper limit or a confidence interval:
she/he simply applies the chosen procedure to construct a
confidence interval $(s_1,s_2)$. If the lower bound turns out to be
strictly equal to zero: $s_1=0$, then an upper limit is quoted, and if
the lower bound is positive: $s_1>0$, then a confidence interval is quoted.    
This versatile procedure, however, has one subtle problem. The fact
that the constructed 90\% confidence interval has a 
non-zero lower bound does not guarantee that a signal of high
statistical significance is observed. There is nothing surprising
about that. A measured signal rate is usually quoted if the
statistical significance of the signal 
exceeds 3 which corresponds to 99.87\% of the area under a Gaussian
peak. At the same time, typical confidence levels used to quote upper
limits are 90\% and 95\%. For example, at $b=1$ and $n=3$
the 90\% confidence interval obtained by the procedure of
Ref.~\cite{feldman} equals $(0.10,6.42)$. At the same time, the
statistical significance $\sigma$ calculated as $1/\sqrt{2\pi} 
\int_{\sigma}^{\infty} e^{-{x^2}/{2}} dx =
\sum_{k=n}^{\infty} e^{-b} b^k/k!$ is only 1.4. 
Thus, the non-zero lower bound has no clear interpretation: it is a
vague indication that a clear signal might be observed in a future
experiment. On the contrary, if one chooses to construct a one-sided
interval and thus quote an upper limit, this clearly expresses the fact that a
signal of sufficient statistical significance was not observed and
therefore the signal rate is believed to be consistent with zero.

%One cannot just quote
%the upper bound 6.42 as a 90\% upper limit since this clearly would be
%an overcoverage. The most consistent way to deal with this situation
%would be to quote results of {\em all} experiments in the form of
%confidence intervals obtained by the technique of
%Ref.~\cite{feldman}. 

%But this is hardly an option since many
%physicists have ideas of their own. In this situation a non-zero lower
%bound $s_1$ becomes a redundant number which has no clear
%interpretation. It is there, but it is irrelevant as long as you
%compare your result with results of other experiments obtained with
%different upper limit estimation techniques.

\subsection{Unbinned Extended Maximum Likelihood Fit}
\label{sec:lkh}

Unbinned extended maximum likelihood fits~\cite{lyons,barlow} have
become popular in the past few years. This is an excellent analysis tool
for rare signal searches since, unlike a standard maximum likelihood, 
an extended maximum likelihood correctly incorporates the Poisson
error on the number of observed events. Thus, when one observes
contributions from two processes
(e.g., signal and background), one should use an extended unbinned
likelihood:
\begin{equation}
\label{eq:lkh}
\lkh(s,b) = \frac{e^{-(s+b)}}{N!} \prod_{i=1}^{N} (s\ss_i + b\bb_i)
\end{equation}
instead of a standard likelihood:
$$
\lkh(f_s) = \prod_{i=1}^{N} (f_s\ss_i + (1-f_s)\bb_i)\ .
$$
Here $\ss_i$ and
$\bb_i$ represent the 
signal and background spatial pdf's, respectively. $N$ is the total number of
events observed in the signal region and in the vicinity, and $s$ and
$b$ are the signal and
background rates, respectively.

It has also become a common practice to extract an upper limit value
by integrating a likelihood function:
\begin{equation}
\label{eq:int_s}
1 - \alpha = \frac{ \int_{0}^{s_0} \lkh(s)ds }{ \int_{0}^{\infty}
\lkh(s)ds }\ , 
\end{equation}
where $s_0$, as usually, denotes the value of an upper limit. 
The likelihood $\lkh(s)$ is typically obtained by integrating the
two-dimensional likelihood $\lkh(s,b)$ over the parameter space $b$:
\begin{equation}
\label{eq:lkh_int}
\lkh(s) = \int_{0}^{\infty} \lkh(s,b) db\ .
\end{equation}
This technique is applied, for instance, to
estimate upper limits in rare $B$ decay
studies~\cite{rare_b1,rare_b2}. It is important to realize 
that the integration of the likelihood implicitly uses the Bayesian approach.
As it was pointed out by Cousins~\cite{cousins}, {\em in particle physics,
the prior is almost always taken uniform (where non-zero), although
this assumption goes unemphasized by those who merely report that they
``integrated the likelihood function''}. 

To understand this comment, one should recall the definition of a
likelihood function:
\begin{equation}
\label{eq:lkh_def}
\lkh(s,b) = f(\vec{x}|s,b)\ ,
\end{equation}
where $\vec{x}=\left\{N,x_1,x_2,...,x_N\right\}$ is a set of
observables, the quantity $x_i$ ($i=1,...,N$) is the coordinate of the
$i$th event, 
the symbol $f$ denotes the corresponding pdf, and a vertical
line $|$ always implies conditional distribution. Applying Bayes'
Theorem, one obtains:
$$
\lkh(s,b) = \frac{ f(s,b|\vec{x}) f(\vec{x}) }{ f(s,b) }\ .
$$
If signal and background are assumed independent\footnote{The
assumption of their independence seems natural. However,
Prosper~\cite{prosp1} derived a non-informative joint prior $f(s,b)$,
where the signal and background contributions are non-factorizable. In
fact, this assumption is not as obvious as it may seem.}, their joint
pdf must factorize: $f(s,b)=f(s)f(b)$. Multiplying both sides of the
above equation 
by $f(b)$ and integrating over $b$, one obtains:
$$
\int_{0}^{\infty} \lkh(s,b) f(b) db =
\frac{ f(s|\vec{x}) f(\vec{x}) }{ f(s) }\ .
$$
The quantity on the right side of the equation can be
recognized as $\lkh(s)$. Therefore,
\begin{equation}
\label{eq:lkh_b}
\lkh(s) = \int_{0}^{\infty} \lkh(s,b) f(b) db\ .
\end{equation}
In the small signal limit
$\ss_i\ll\bb_i$, the likelihood~(\ref{eq:lkh}) is
expressed as:
$$
\lkh(s,b) = \frac{e^{-(s+b)}}{N!} b^N \prod_{i=1}^{N} \bb_i\ .
$$
Due to the fact that the contributions from the signal and background
rates $s$ and $b$ decouple, the
background pdf $f(b)$ in Eqn.~(\ref{eq:lkh_b}) cannot change the ratio
$\int_{0}^{s_0} \lkh(s)ds / \int_{0}^{\infty} \lkh(s)ds$. 
But when the approximation $\ss_i\ll\bb_i$ is no 
longer valid, this is not so.
Thus, Eqn.~(\ref{eq:lkh_int}) treats the pdf $f(b)$ as prior knowledge
and assumes that it is flat in the interval $(0,+\infty)$.

Furthermore, the posterior distribution $f(s|\vec{x})$ can be
represented as:
$$
f(s|\vec{x}) = \frac{ \lkh(s) f(s) }{ f(\vec{x}) }\ ,
$$
which is similar to the Bayesian formula~(\ref{eq:poster}). Thus,
Eqn.~(\ref{eq:int_s}) implicitly assumes a flat prior $f(s)$ and
integrates $f(s|\vec{x}) \propto \lkh(s)$ to extract an upper
limit. When the experiment is dominated by background, i.e.,
$\ss_i\ll\bb_i$, the 
likelihood~(\ref{eq:lkh_int}) is proportional to $e^{-s}$ and the
extracted value of an upper limit is equal to 2.3 at 90\% confidence
level. The approximation $\ss_i\ll\bb_i$ is equivalent to the
situation when no events are 
observed in the signal region ($n=0$), but the expected number of
background events is non-zero ($b\neq 0$). 
Under these conditions, the Bayesian method~(\ref{eq:pdg96}) with a
flat prior gives the $e^{-s}$ behavior as well.
Neither
the classical approach, nor the Bayesian methods with alternative priors
give the $e^{-s}$ behavior for $n=0$ and $b\neq 0$; all these
techniques give smaller upper limit values. Thus, the integration of
the likelihood
and the Bayesian method with a flat prior
give the most conservative estimates in background-dominated analyses.

One can easily avoid the implicit use of the Bayesian method and set
an upper limit using the frequentist definition. It is straightforward
to implement this approach by running a Monte Carlo job. First,
the observed data
$\vec{x}_{obs}$ is fitted to the likelihood function~(\ref{eq:lkh}) to
extract estimates $s_{obs}$ and $b_{obs}$ of the true signal and
background rates. 
Then, for every assumed value of an upper limit $s_0$, a Monte Carlo
sample, consisting of a large number of experiments, is generated.
For each experiment the
number of signal and background events are generated assuming Poisson
distributions:
$$
f(s) = e^{-s_0} \frac{ s_0^s }{s!};\ \ \
f(b) = e^{-b_{obs}} \frac{ b_{obs}^b }{b!};\ \ \ N=s+b;
$$
and the coordinates of events are generated, based on the spatial pdf's
$\ss_i$ and $\bb_i$. Then the outcome of every experiment is fitted to
the same likelihood function~(\ref{eq:lkh}) and the
distribution of measured signal rates $f(s_{meas})$ is plotted.
The confidence level corresponding to this value of $s_0$ is estimated
as: 
\begin{equation}
\label{eq:ul_mc}
1 - \alpha = \frac{ \int_{s_{obs}}^{\infty} f(s_{meas}) ds_{meas} }
{ \int_0^{\infty} f(s_{meas}) ds_{meas} }\ .
\end{equation}
Thus, a value of $s_0$ is chosen that gives the
required confidence level $(1-\alpha)$. This agrees with the
frequentist approach when an experimentalist draws inference about an
unknown parameter on the basis of observed data only, without
making any subjective assumption.

%Then a Monte Carlo sample, consisting of
%a large number of experiments, is generated. For each experiment the
%number of signal and background events are generated assuming Poisson
%distributions:
%$$
%f(s) = e^{-s_{obs}} \frac{ s_{obs}^s }{s!};\ \ \
%f(b) = e^{-b_{obs}} \frac{ b_{obs}^b }{b!};\ \ \ N=s+b;
%$$
%and the coordinates of events are generated, based on the spatial pdf's
%$\ss_i$ and $\bb_i$. Then the outcome of every experiment is fitted to
%the same likelihood function~(\ref{eq:lkh}) and the
%distribution of measured signal rates $f(s_{meas})$ is plotted. The
%last step is to extract the value $s_0$ of an upper limit from the equation:
%$$
%1 - \alpha = \frac{ \int_0^{s_0} f(s_{meas}) ds_{meas} }
%{ \int_0^{\infty} f(s_{meas}) ds_{meas} }\ .
%$$
%The extracted value $s_0$ of an upper limit does not rely on any
%assumption on the signal and background prior pdf's $f(s)$ and
%$f(b)$. This agrees with the conventional approach when an
%experimentalist draws inference about an unknown parameter on the
%basis of the observed data only, without making any subjective assumption.

\section{Calculation of Upper Limits, Based on Statistical
Significance of a Signal}
\label{sec:ss_gen}

The Bayesian and classical approaches described in
Sections~\ref{sec:bayes} and \ref{sec:class} are based on the
conditional probability $f(n|s)$, i.e., the probability to observe $n$
events in the signal region, given the signal rate $s$. 
However, the number
of events observed in the signal region by itself is not so important,
since the usual goal of an experiment is not to 
count events in the signal region, but to observe a
signal of high statistical significance. These approaches do not take
into account the fact that the 
number of background events fluctuates too. 
For example,
the classical method~(\ref{eq:class_ul}) estimates the confidence
level as a probability to observe a larger number of events in the
signal region than the number observed in the experiment, given the
assumed value $s$ of the signal rate. 
%But at the same time the number of background events in the sideband
%can fluctuate up to high values, which means that an experimentalist
%does not observe any signal, or it can fluctuate down to
%zero, which means that an experimentalist does observe a very clean
%signal.
But, at the same time, the number of background events in the sideband
can fluctuate up to high values or down to zero. The former implies
that no signal is observed, and the latter implies that a very clean
signal is observed.
The methods based on the pdf $f(n|s)$ take into account both these
possibilities and thus do not distinguish between signals of high and
low statistical significance.

Therefore, one should replace the conditional pdf $f(n|s)$ with the
conditional pdf $f(\sigma|s)$, where $\sigma$ represents the
statistical significance of a signal. Before we proceed to derivation
of the related mathematical formalism, we need to discuss various
definitions of statistical significance.

\subsection{Definition of Statistical Significance}
\label{sec:comment}

If $n$ events are observed in the signal region and $b$ is the
estimated background rate, then the statistical significance
$\sigma$ of a signal is defined by
\begin{equation}
\label{eq:ss_prob}
\frac{1}{\sqrt{2\pi}} 
\int_{\sigma}^{\infty} e^{-{x^2}/{2}} dx =
\sum_{k=n}^{\infty} e^{-b} \frac{b^k}{k!}\ ,
\end{equation}
that is, it represents the probability 
of observing $n$ or a larger number of background events in an
identical independent experiment.
%for background to fluctuate up
%to the observed number of events and higher. 
The background rate $b$
can be estimated in a number of ways. If the background rate is
estimated {\em independently} of the data seen, e.g., from a Monte
Carlo analysis or from another data sample that is known to contain no
signal events, then it is common to assume that the estimated
background rate is proportional to the rate observed in the
independent experiment:
\begin{equation}
\label{eq:b_ind}
b = \zeta_{ind} n_{ind}\ ,
\end{equation}
where $n_{ind}$ is the number of events observed in the independent
experiment, $\zeta_{ind}$ is the corresponding scale factor and the
subscript $ind$ implies that the background rate is estimated
independently of the data seen. If the background rate is estimated
{\em from data}, then the situation is more complicated. In this case,
the background rate is usually estimated from sideband, i.e., a region
which is located near the signal region and which contains no signal events:
\begin{equation}
\label{eq:b_stand}
b = \zeta_{sb} n_{sb}\ .
\end{equation}
Here $n_{sb}$ is the number of events observed in the
sideband, and $\zeta_{sb}$ is the sideband-to-signal scale factor.
Under the assumption of flat background, the scale factor is given by:
\begin{equation}
\zeta_{sb} = \frac{A_{sig}}{A_{sb}}\ ,
\end{equation}
where $A_{sig}$ is the area of the signal region, and $A_{sb}$ is the
area of the sideband.

The definition~(\ref{eq:b_stand}) of the background rate is, in my
opinion, incorrect. 
It works only if the
sideband is much larger than the signal region, and thus the estimate
of the expected background rate $b$ is accurate enough. However, in a
situation when the areas of signal and sideband regions are
comparable, Eqn.~(\ref{eq:b_stand}), combined with
Eqn.~(\ref{eq:ss_prob}), overestimates the significance of 
a signal for large numbers of observed events $n>b$, increasing the
probability of a ``discovery''. This is caused by the fact that the
formulas~(\ref{eq:ss_prob}) and (\ref{eq:b_stand}) answer the wrong
question. The correct question is: what is the
probability of observing $n$ or a larger number of events in
the signal region 
{\it if the true signal rate is zero}? To answer this question, one
has to assume that {\it all} events, that were observed in the
experiment, came from background, and therefore one has to redefine:
\begin{equation}
\label{eq:b_my}
b = \zeta N\ ,
\end{equation}
where $N$ is the number of events observed in the entire region, which
includes the signal region, sideband and,
perhaps, an intermediate region between the signal and sideband
regions, and $\zeta$ is the corresponding scale factor. Under the
assumption of flat background, the scale factor is given by:
\begin{equation}
\zeta = \frac{A_{sig}}{A}\ ,
\end{equation}
where $A$ is the area of the entire region.
In the same spirit, statistical significance, as defined via likelihood, is
given by:
\begin{equation}
\sigma = \sqrt{ -2\, ln\, \lkh(0)/\lkh_{max} }\ ,
\end{equation}
i.e., it gives a number of standard deviations from the observed signal
rate, which maximizes $\lkh(s)$, to the zero signal rate, under the
assumption that $-2\, ln\, \lkh(0)/\lkh_{max}$ is distributed as $\chi_1^2$.
Formula~(\ref{eq:b_my}) should be used only to estimate statistical
significance of a signal; for upper limit calculation one is not
allowed to assume that all observed events come from background and
one has to apply the standard definition(\ref{eq:b_stand}) of an
expected background rate.

A numerical discrepancy between the definitions~(\ref{eq:b_stand}) and
(\ref{eq:b_my}) can be illustrated on the following hypothetical
example. Let us assume that the background spatial pdf is flat in the
vicinity of the signal region,
that the area of the sideband is equal to that
of the signal region, and that one event is observed in
the signal region and no events are observed in the sideband. Then an
experimentalist, who uses the definition~(\ref{eq:b_stand}) of the
statistical significance, would claim that she/he observes a very clean
signal ($\sigma=+\infty$), while an experimentalist, who uses the
definition~(\ref{eq:b_my}), will estimate the statistical significance of
the observed signal as $\sigma=0.27$. The latter reflects the fact
that, due to the specific 
choice of the sideband and signal regions, the observed statistic (one
event) is insufficient for positive identification of the signal.

This example is purely hypothetical. Of course, in the situation
when only one event is observed and the area of the sideband is
comparable to that of the signal region, most of experimentalists
would choose to quote an upper limit instead of a measurement. A more
realistic situation occurs when the sideband is somewhat larger than
the signal region, there are some events both in the signal region and
in the sideband and the calculated statistical significance is close
to three. Then one has to make a binary decision: if the statistical
significance is larger than three, then a measurement is quoted, otherwise
an upper limit is quoted. In this situation the choice of a specific
procedure becomes fairly important, and the example above shows that
the numerical discrepancy between the two approaches can be significant.

The conclusion that I would like to reach in this Section is that
there are two situations that should be treated differently. The
first situation occurs when the background rate is estimated
independently of the data seen. In this case one has to adopt the
definition~(\ref{eq:b_ind}) of the background rate and therefore
define statistical significance as:
\begin{equation}
\label{eq:ss_ind}
\frac{1}{\sqrt{2\pi}} 
\int_{\sigma}^{\infty} e^{-{x^2}/{2}} dx =
\sum_{k=n}^{\infty} e^{-\zeta_{ind}n_{ind}}
\frac{(\zeta_{ind}n_{ind})^k}{k!}\ . 
\end{equation}
The second situation takes place when the background rate is estimated
from the same data sample which is used to draw inference about the
unknown signal rate. As shown above, in this case one has to adopt the
definition~(\ref{eq:b_my}). Statistical significance is then defined
as:
\begin{equation}
\label{eq:ss_my}
\frac{1}{\sqrt{2\pi}} 
\int_{\sigma}^{\infty} e^{-{x^2}/{2}} dx =
\sum_{k=n}^{\infty} e^{-\zeta N}
\frac{(\zeta N)^k}{k!}\ . 
\end{equation}
These two formulas will be used to estimate upper limits, based on the
statistical significance of a signal.

\subsection{Upper Limit Calculation}
\label{sec:ss}

As shown in Section~\ref{sec:comment}, the statistical significance $\sigma$
of a signal is defined either by Eqn.~(\ref{eq:ss_ind}) or by
Eqn.~(\ref{eq:ss_my}). I will take Eqn.~(\ref{eq:ss_my}) as an
example and proceed to derive a {\em cumulative density function}
(cdf) $P(\sigma\leq\sigma')$. If Eqn.~(\ref{eq:ss_ind}) is chosen, the
derivation goes through similar steps; thus, only the final result
will be quoted. 

When the background and signal rates are estimated from the same data
sample, one can rewrite Eqn.~(\ref{eq:ss_my}) as:
\begin{equation}
\label{eq:ss_use}
\frac{1}{\sqrt{2\pi}} 
\int_{\sigma}^{\infty} e^{-{x^2}/{2}} dx =
\sum_{k=n}^{\infty} e^{-\zeta(n+n_{out})}
\frac{\left[\zeta(n+n_{out})\right]^k}{k!}\ ,
\end{equation}
where $n$ is the number of events observed in the signal region,
$n_{out}$ is the number of events observed in the outer region, which
covers the sideband and, perhaps, an intermediate region between the
signal region and the sideband, and $\zeta$ is the corresponding scale
factor. Under the assumption of flat background, the scale factor is
given by:
\begin{equation}
\zeta = \frac{A_{sig}}{A_{sig}+A_{out}}\ ,
\end{equation}
where $A_{sig}$ and $A_{out}$ are the areas of the signal and outer
regions respectively.
The random variables $n$ and $n_{out}$ are
drawn from independent Poisson distributions:
\begin{equation}
\label{eq:dists}
n \sim \mbox{Poisson}(s+b);\ \ \ n_{out} \sim \mbox{Poisson}(\lambda_{out});
\end{equation}
where $\lambda_{out}$ is the mean of the Poisson distribution that
controls the number of events in the outer region $A_{out}$. The best
unbiased estimator of $\lambda_{out}$ is the number of events actually
observed outside the signal region:
\begin{equation}
\hat{\lambda}_{out} = n_{out,obs}\ ,
\end{equation}
and thus the equality $\lambda_{out} = n_{out,obs}$ is implied in the
further discussion.
In Eqn.~(\ref{eq:dists}) the expected number of background events $b$
in the signal region is estimated in the traditional way:
$b=\zeta_{sb}n_{sb,obs}$, and its value is not used to
determine the statistical significance of the signal. 

The cdf $P(\sigma\leq\sigma')$ cannot be expressed in a convenient
analytical form.
However, the problem can be simplified by introducing a new variable:
\begin{equation}
\label{eq:p}
p = \sum_{k=0}^{n-1} e^{-\zeta(n+n_{out})}
\frac{\left[\zeta(n+n_{out})\right]^k}{k!}\ .
\end{equation}
Hence,
$$\frac{1}{\sqrt{2\pi}} 
\int_{\sigma}^{\infty} e^{-{x^2}/{2}} dx = 1 - p\ ;\ \ \ 0\leq p\leq 1\ .$$
The variable $p$ is a monotone function of $\sigma$. Therefore, its
cdf can be obtained by a one-to-one transformation:
\begin{equation}
\label{eq:p_and_sigma}
P( p(\sigma)\leq p(\sigma') ) = P( \sigma\leq\sigma' )\ ,
\end{equation}
and one can use the variable $p$ instead of $\sigma$ to set an upper
limit. The cdf of $p$ is given by:
%\begin{equation}
%\label{eq:cdf_p}
%P(p\leq p') = \sum_{n=0}^{\infty} \sum_{n_{out}=n'_{out}}^{\infty}
%e^{-(s+b)} \frac{(s+b)^n}{n!} e^{-\lambda_{out}}
%\frac{\lambda_{out}^{n_{out}}}{n_{out}!}\ ,
%\end{equation}
%where $n'_{out}=n'_{out}(n)$ is the smallest non-negative integer
%which satisfies the inequality
%\begin{equation}
%\label{eq:n_out}
%\sum_{k=0}^{n-1} e^{-\zeta(n+n'_{out})}
%\frac{\left[\zeta(n+n'_{out})\right]^k}{k!} \leq p'
%\end{equation}
%for the given value of $n$. Rearranging terms, one obtains:
\begin{equation}
\label{eq:1-p}
1 - P(p\leq p'|s) = \sum_{n=1}^{\infty} \sum_{n_{out}=0}^{n'_{out}-1}
e^{-(s+b)} \frac{(s+b)^n}{n!} e^{-\lambda_{out}}
\frac{\lambda_{out}^{n_{out}}}{n_{out}!}\ ,
\end{equation}
where $n'_{out}=n'_{out}(n)$ is the smallest non-negative integer
which satisfies the inequality:
\begin{equation}
\label{eq:n_out}
\sum_{k=0}^{n-1} e^{-\zeta(n+n'_{out})}
\frac{\left[\zeta(n+n'_{out})\right]^k}{k!} \leq p'
\end{equation}
for the given value of $n$.

The infinite sum over $n$ in Eqn.~(\ref{eq:1-p}) converges quickly
and can be easily calculated numerically. In a simulation, described
in Section~\ref{sec:toy_model}, the summation over $n$ from 1
to 1000 was enough to achieve an accuracy of $10^{-6}$ or better
for $(1-P(p\leq p'|s))$.

Now one can choose the approach one would like to use. All the
techniques described in Sections~\ref{sec:bayes} and \ref{sec:class}
are applicable, but now instead of a conditional pdf $f(n|s)$ one
would have to use the conditional cdf $P(p\leq p'|s)$. The easiest
thing to try is to employ the classical approach and define the
confidence interval as $(0,s_0)$. In this case, through the reasoning
described in Section~\ref{sec:class}, one arrives at the formula
similar to Eqn.~(\ref{eq:class_ul}):
\begin{equation}
\label{eq:ul_p}
1 - \alpha = 1 - P(p\leq p_{obs}|s_0)\ ,
\end{equation}
where
\begin{equation}
\label{eq:p_obs}
p_{obs} = \sum_{k=0}^{n_{obs}-1} e^{-\zeta(n_{obs}+n_{out,obs})}
\frac{\left[\zeta(n_{obs}+n_{out,obs})\right]^k}{k!}
\end{equation}
is the observed value of the variable $p$.

When no events are observed in the signal region ($n_{obs}=0$),
one obtains $p_{obs}=0$, therefore $n'_{out}=+\infty$ for any $n\geq
1$, and Eqn.~(\ref{eq:ul_p}) is reduced to:
$$1 - \alpha = 1 - e^{-(s_0+b)}\ , $$
which coincides with the classical
expression~(\ref{eq:class_ul}). This reflects the fact that in this
case background fluctuations are irrelevant as the statistical
significance $\sigma$ is always equal to $-\infty$, no matter how many
background events we observe in the sideband.

When no events are observed outside of the signal region
($\lambda_{out}=n_{out,obs}=0$ and therefore $b=0$), one obtains
$n'_{out}\geq 1$ for any $n\geq 1$ and $p_{obs}>0$, therefore
%$\sum_{n_{out}=0}^{n'_{out}-1} e^{-\lambda_{out}}
%\lambda_{out}^{n_{out}}/n_{out}!$ 
the sum over $n_{out}$ in Eqn.~(\ref{eq:1-p})
is always equal to 1, and the
formula~(\ref{eq:ul_p}) is reduced to:
$$1 - \alpha = 1 - e^{-s_0}\ ,$$
which again is identical to the classical
expression~(\ref{eq:class_ul}). This reflects the fact that no
background fluctuations are expected in this case, and hence, the statistical
significance $\sigma$ is a function of the signal rate $s$ only.

If the background rate is estimated independently of the data seen,
then one should start from the definition~(\ref{eq:ss_ind}) of
statistical significance and repeat the same logical steps to arrive
at an equation similar to Eqn.~(\ref{eq:1-p}). 
The random variables $n$ and $n_{ind}$ are
drawn from independent Poisson distributions:
\begin{equation}
n \sim \mbox{Poisson}(s+b);\ \ \ n_{ind} \sim \mbox{Poisson}(\lambda_{ind});
\end{equation}
where $\lambda_{ind}$ is the mean of the Poisson distribution that
controls the number of events in an independent sample which is used
to estimate the background rate. The best
unbiased estimator of $\lambda_{ind}$ is the number of events actually
observed in the independent data sample:
\begin{equation}
\hat{\lambda}_{ind} = n_{ind,obs}\ ,
\end{equation}
and thus the equality $\lambda_{ind} = n_{ind,obs}$ is implied. Now
the background rate used in the definition of statistical significance
and the background rate used to calculate an upper limit are
estimated similarly. Without losing generality, one can stipulate
that:
\begin{equation}
\lambda_{ind} = b/\zeta_{ind}\ .
\end{equation}
The cdf of $p$ is now given by:
\begin{equation}
1 - P(p\leq p'|s) = \sum_{n=1}^{\infty} \sum_{n_{ind}=0}^{n'_{ind}-1}
e^{-(s+b)} \frac{(s+b)^n}{n!} e^{-b/\zeta_{ind}}
\frac{(b/\zeta_{ind})^{n_{ind}}}{n_{ind}!}\ ,
\end{equation}
where $n'_{ind}=n'_{ind}(n)$ is the smallest non-negative integer
which satisfies the inequality
\begin{equation}
\sum_{k=0}^{n-1} e^{-\zeta_{ind}n'_{ind}}
\frac{(\zeta_{ind}n'_{ind})^k}{k!} \leq p'
\end{equation}
for the given value of $n$.

\section{Comparison of Various Approaches Using a Toy Model}
\label{sec:toy_model}

The performance of all approaches discussed in this paper is compared
using the following toy model. The total observation region is defined
as an interval $(-10,10)$. 
The signal spatial pdf is taken to be a Gaussian with zero mean and unit
variance, 
and the background spatial pdf is taken to be flat.
Under these conditions, the signal
region is defined as $(-2.5,2.5)$, and the sidebands are defined as
$(-10,-5)$ and $(5,10)$. 

The value of the expected background rate is taken to be $b=0, 1, ..., 5$
consecutively, which corresponds to the number of events observed in
the sidebands $n_{sb}=0, 2, ..., 10$. The number of events in the
intermediate region $(-5,-2.5)$ and $(2.5,5)$ is set equal to the
expected number $b$ of background events in the signal region, since
the area of 
the signal region is equal to that of the intermediate
region. Positions of events in the sideband and intermediate regions
are generated under the assumption of a uniform spatial pdf
$\bb_i$. For every value of the expected background rate $b$, upper
limits obtained with various approaches are estimated for the number
$n$ of events in the signal region 
varying in integer steps from 0 to 6. The results are plotted in
Fig.~\ref{fig:all}. Inside the signal region all events are positioned
precisely at zero. It is assumed that the background rate in this
experiment is estimated from sidebands; thus, to implement the
classical method of Section~\ref{sec:ss}, formulas~(\ref{eq:ul_p}) and
(\ref{eq:p_obs}) are used. 
To estimate a 90\% CL upper limit by the Monte
Carlo technique of Section~\ref{sec:lkh}, I find the value $s_0$ of an upper
limit by using the method of binary division. For every assumed value
of $s_0$, I generate a Monte Carlo sample consisting of 50,000
experiments and use the formula~(\ref{eq:ul_mc}) to estimate the
corresponding confidence level. This procedure is repeated until the
value of $s_0$ that gives a 90\% confidence level is obtained. The
required accuracy for the computation of $s_0$ is taken $10^{-2}$. 
%To estimate a 90\% CL upper limit by the Monte
%Carlo technique of Section~\ref{sec:lkh}, I
%generate a Monte Carlo sample consisting of 50,000 experiments for
%each value of $n$ and $b$, and 
%then I pick up the value $s_0$ of an upper limit which is greater than
%the measured signal rate in 45,000 experiments and smaller than the
%measured signal rate in the other 5,000 experiments. 
For comparison,
lower and upper bounds of 90\% confidence intervals obtained by the
procedure~\cite{feldman} are shown with crosses. One cross for a
specific value of $n$ corresponds to the situation when the lower
bound is strictly zero.

%\duplea{htbp}{8b0_freq.eps}{8b1_freq.eps}
%\dupleb{htbp}{8b2_freq.eps}{8b3_freq.eps}
%\duplex{htbp}{8b4_freq.eps}{8b5_freq.eps}
\sixplex{htbp}
{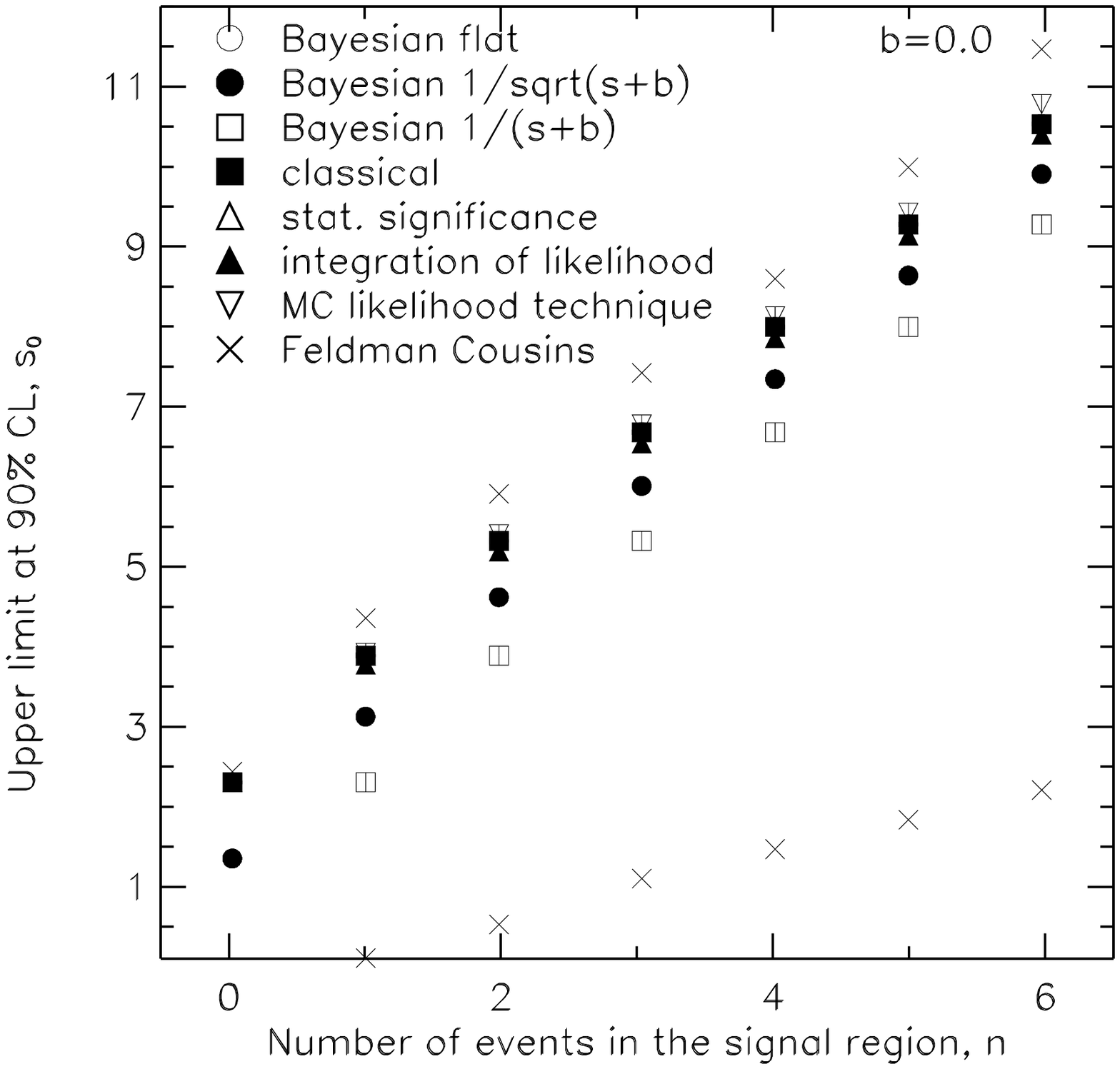}{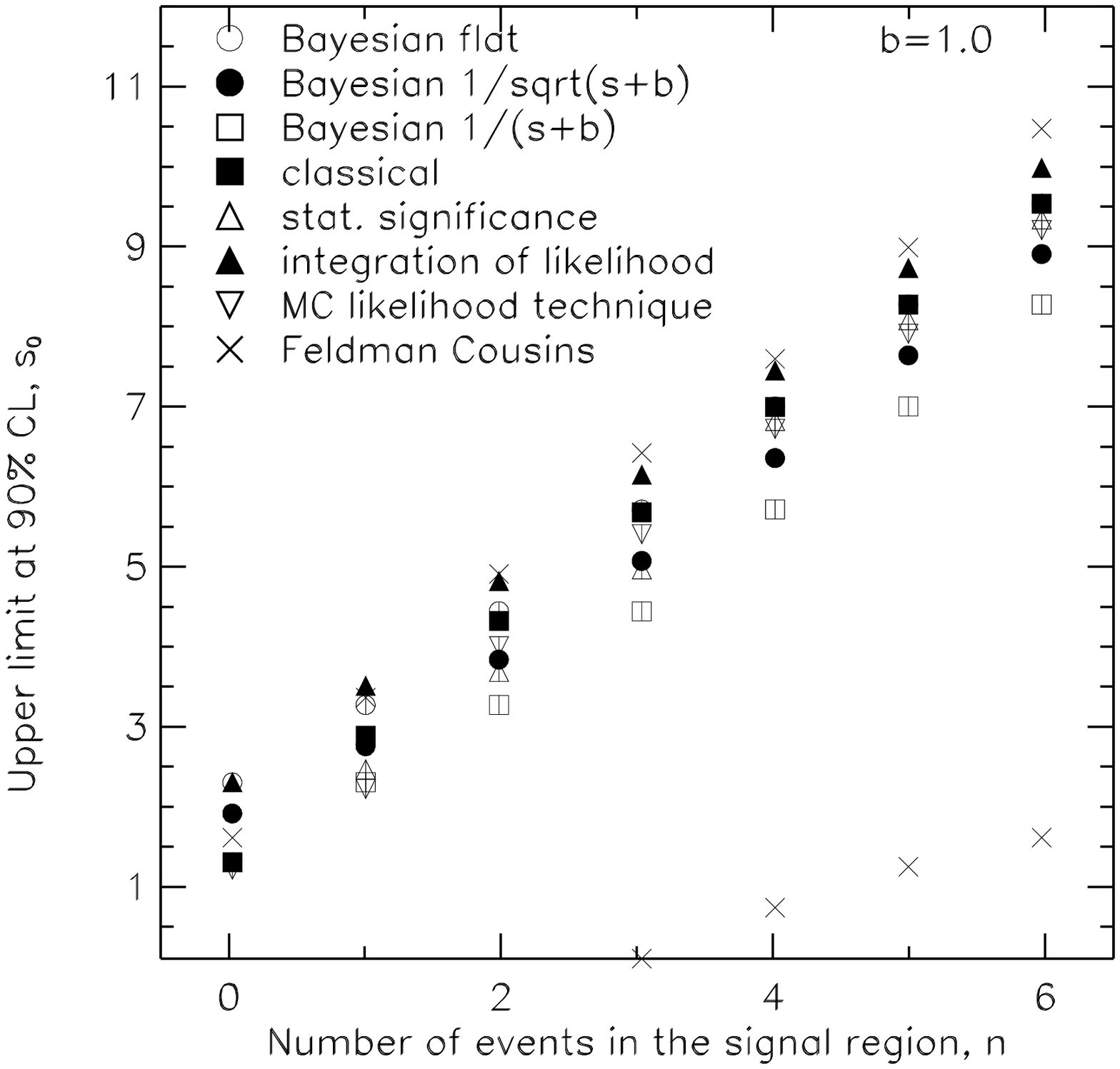}
{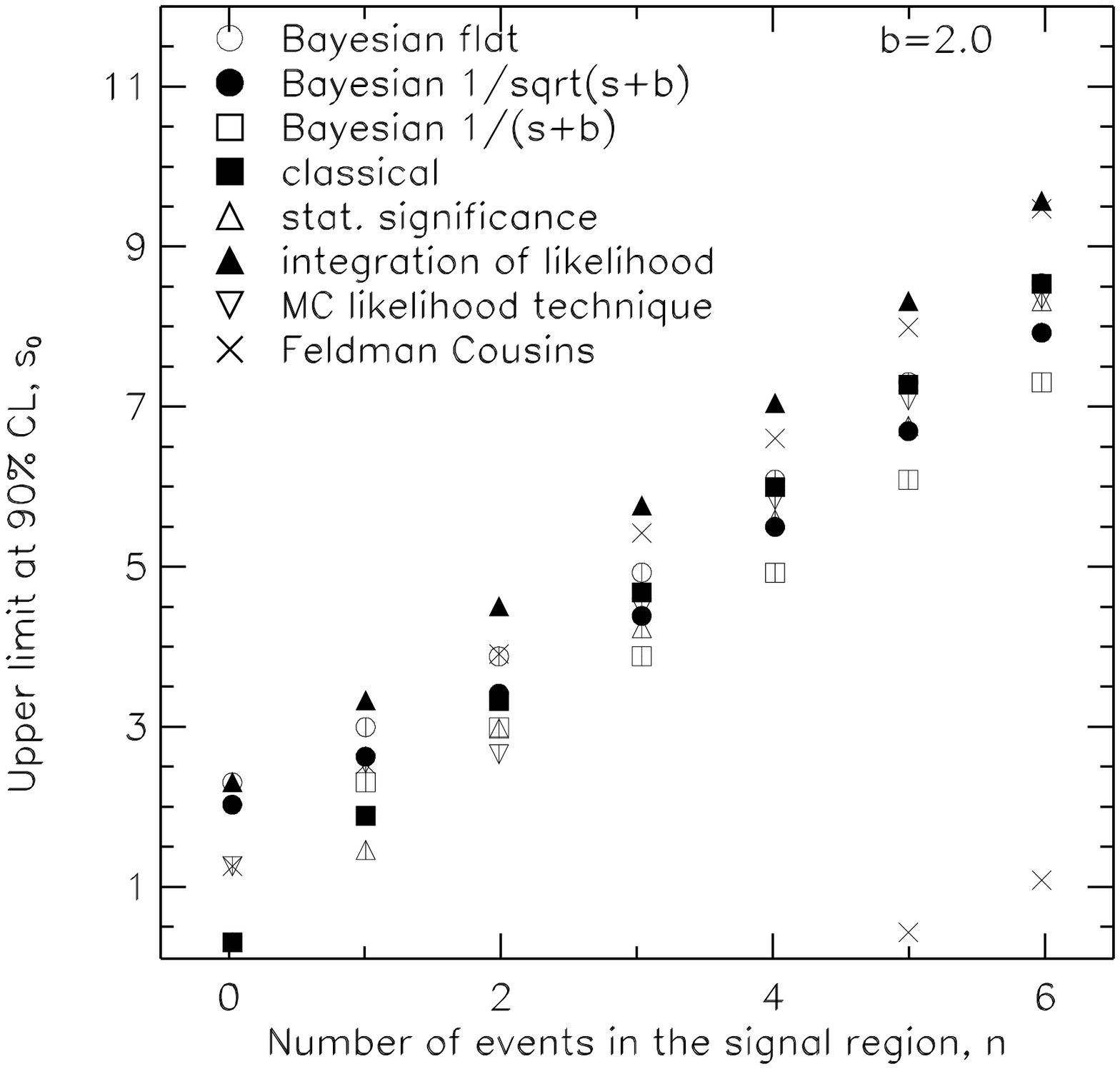}{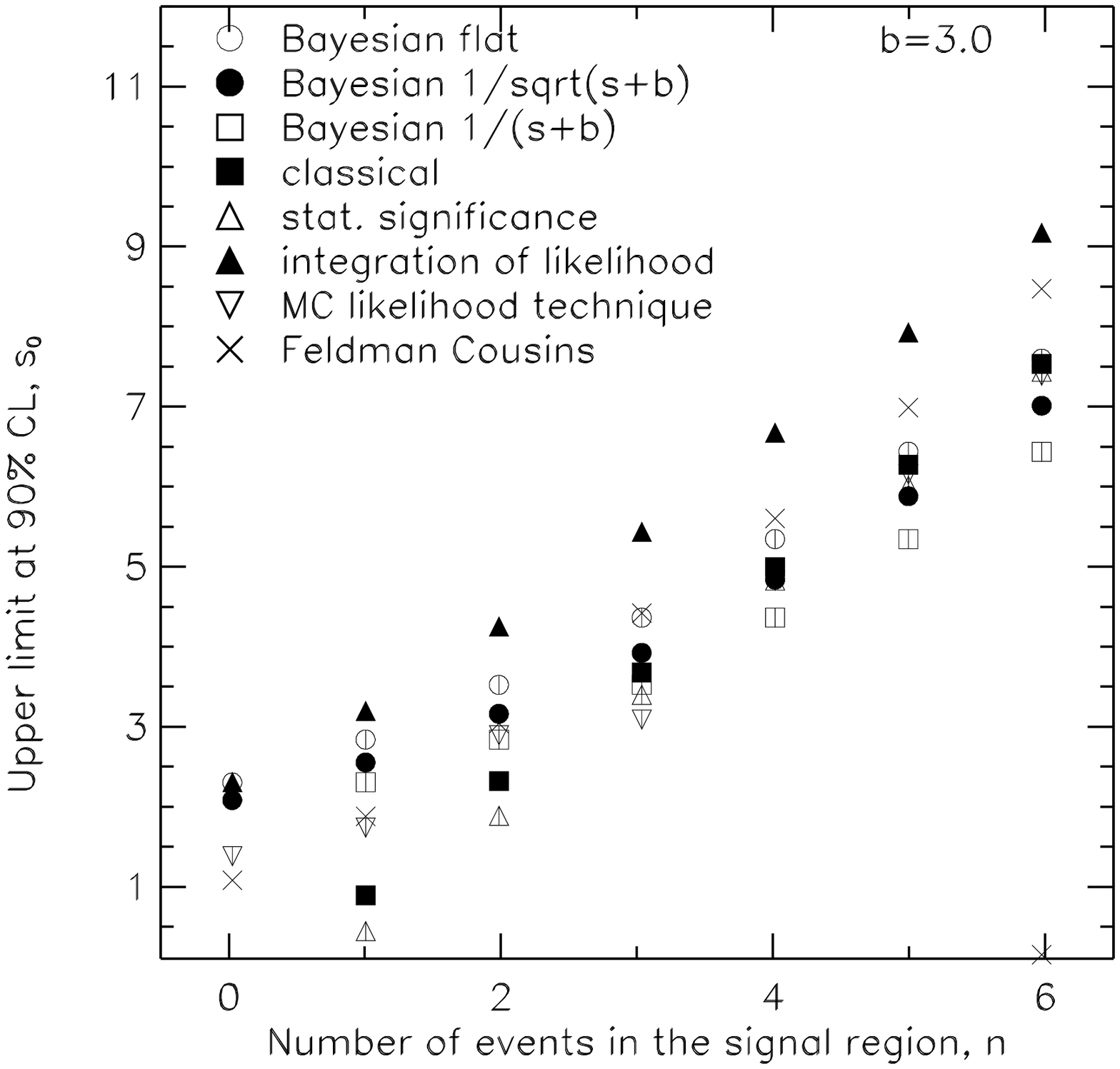}
{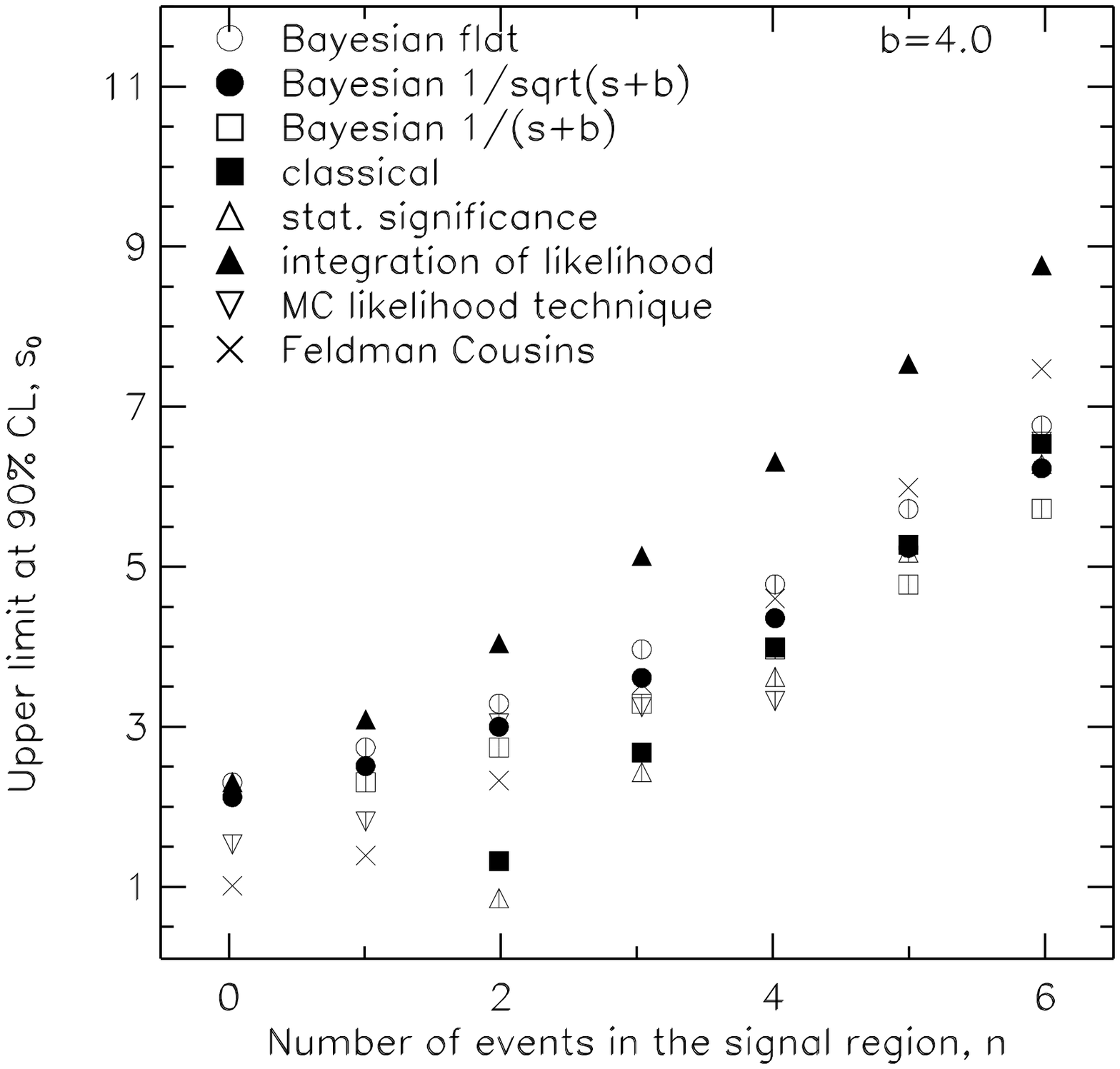}{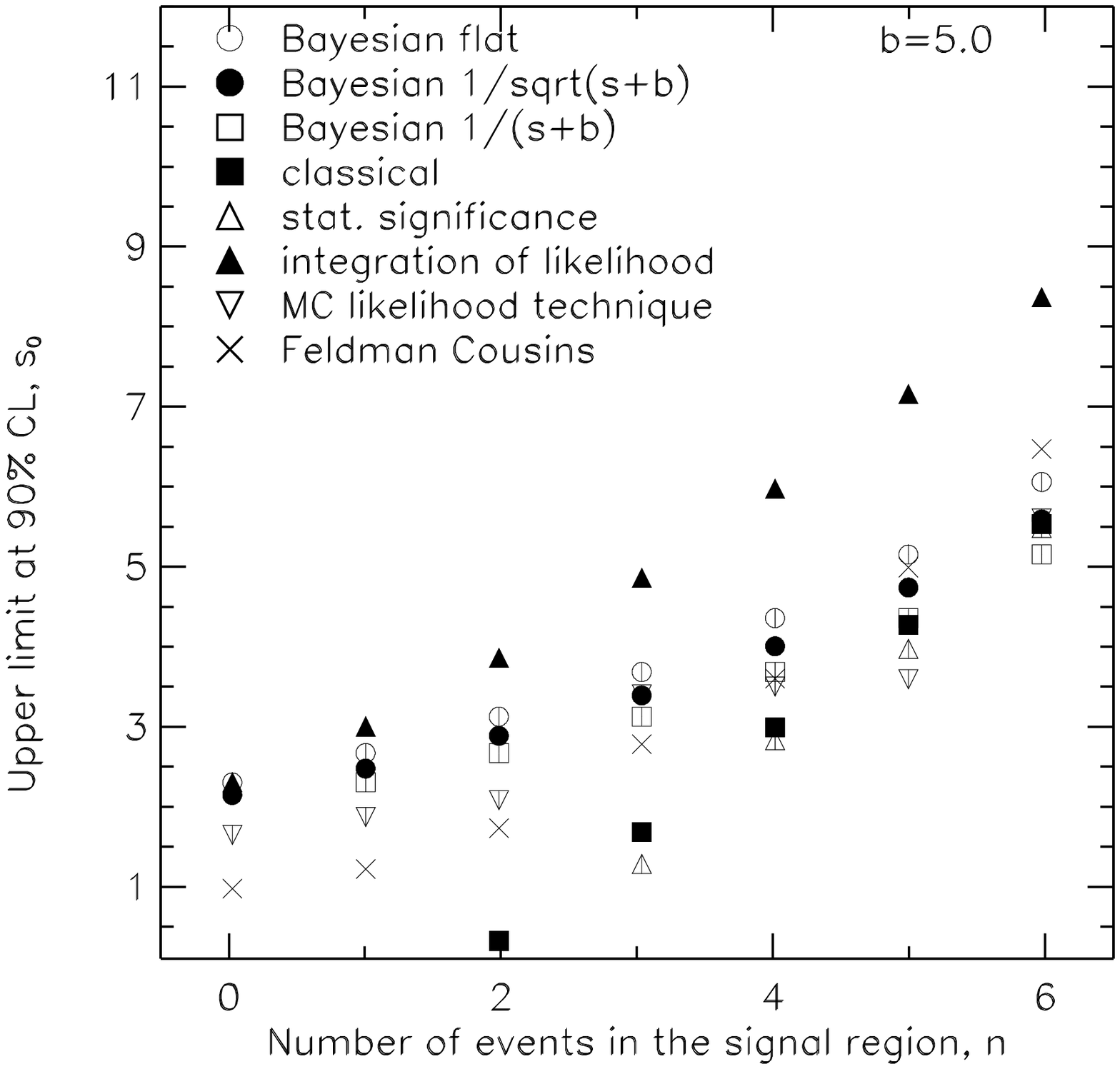}
{all}{Upper limits as functions of the number $n$ of observed events in
the signal region under
the assumptions described in the text. The expected number of
background events in the signal region takes values of 1) $b=0$; 2) $b=1$; 3)
$b=2$; 4) $b=3$; 5) $b=4$; 6) $b=5$.}

As shown in Fig.~\ref{fig:all}, at $n=0$ the Bayesian
method~(\ref{eq:pdg96}) with a flat prior pdf 
and the integration of the likelihood, described in Section~\ref{sec:lkh},
always give 2.30, while 
all the other approaches give somewhat smaller values.
For this particular model, 
the likelihood integration technique of Section~\ref{sec:lkh} turns out to be
the most conservative approach, except for $b=0$. The Bayesian
method with a flat prior pdf always gives larger upper limit values
than the Bayesian methods with the $1/\sqrt{s+b}$ and $1/(s+b)$ priors
and both classical 
approaches discussed in Sections~\ref{sec:class} and \ref{sec:ss}. 
The results produced by the Bayesian method with the $1/(s+b)$ prior
can be obtained by shifting the corresponding results obtained with a
flat prior one step to the right, e.g., $\{n=0,m=0\}$ and
$\{n=1,m=1\}$ obviously produce the same result when substituted into
the gamma-function $\Gamma(n-m+1,s_0+b)$. The
classical approaches of Sections~\ref{sec:class} and 
\ref{sec:ss}
%as well as the Monte Carlo likelihood technique of Section~\ref{sec:lkh},
fail
to set an upper limit at $n\ll b$, i.e., when background dominates over
signal. The classical procedure, based on the statistical significance
of a signal, always gives a smaller upper limit value, compared to the
standard classical approach of Section~\ref{sec:class}, except at
$n=0$. In the specific situation, when the expected 
background is zero, the Bayesian method with a flat prior pdf, both classical
approaches and both likelihood techniques of
Section~\ref{sec:lkh} give identical results.

\section{An Example of a CLEO Analysis: $\tau\to\mu\gamma$}
\label{sec:cleo}

In 1996 the CLEO Collaboration searched~\cite{mugamma} for the
neutrinoless decay
$\tau\to\mu\gamma$ and set an upper limit, which is, so far, the most
stringent limit on the $\tau\to\mu\gamma$ branching fraction. In this
analysis 3 events were observed in the signal region and the expected
number of background events was estimated to be 5.5. 
The signal Monte Carlo and data energy-vs-mass distributions are shown in
Fig.~\ref{fig:mugamma}. The
Bayesian approach with a flat prior pdf was used and the upper limit
value was estimated as 3.6 at 90\% confidence level. 
This value was divided by the integrated luminosity and efficiency
factor and thus an upper limit of $3.0\times 10^{-6}$ at 90\%
confidence level was obtained for the
$\tau\to\mu\gamma$ branching fraction.
%Dividing this number by the
%integrated luminosity 
%and efficiency factor, CLEO obtained the upper limit on the
%$\tau\to\mu\gamma$ branching fraction as $3.0\times 10^{-6}$ at 90\%
%CL. 
In Table~1 are shown the values of upper limits for this analysis
calculated with the alternative techniques.

\simplex{htbp}{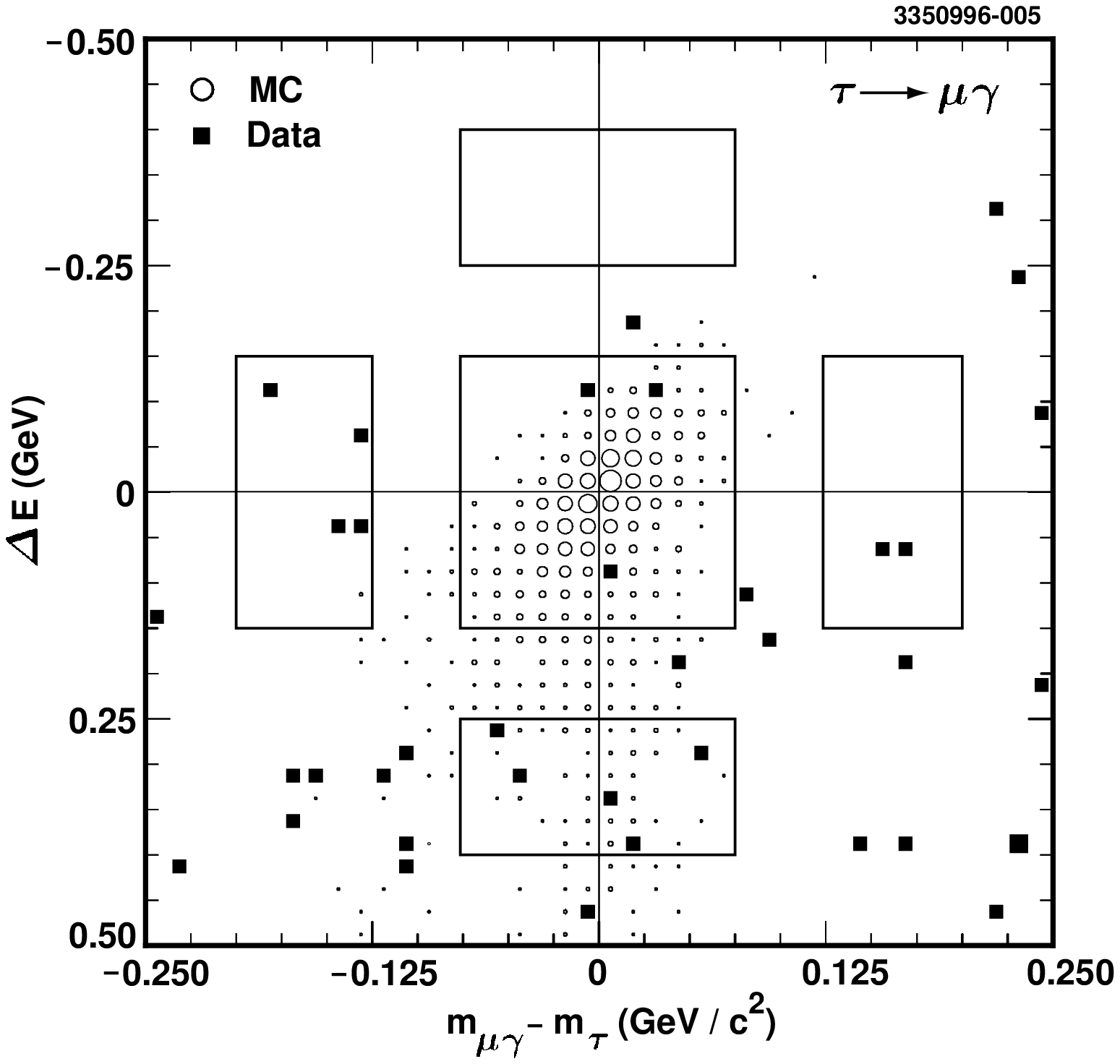}{mugamma}
{$(E_{\mu\gamma}-E_{beam})$ vs $(m_{\mu\gamma}-m_{\tau})$ distribution.
Solid squares represent the data, open circles
represent the signal Monte Carlo distribution.}

%\begin{center}
\begin{table}[bthp]
\caption{}{\bf Upper limits at 90\% CL for the $\tau\to\mu\gamma$ analysis
of Ref.~\cite{mugamma}.}
\begin{tabular}{|c|c|}\hline
{\bf Method}                  	& {\bf Upper limit at 90\% CL} \\ \hline
Bayesian with flat prior   	& {\bf 3.57}    \\ \hline
Bayesian $1/\sqrt{s+b}$      	& 3.30          \\ \hline
Bayesian $1/(s+b)$		& 3.06		\\ \hline
classical               	& 1.18          \\ \hline
classical, based on statistical significance & 1.03 \\ \hline
integration of likelihood	& 2.30		\\ \hline
%MC likelihood technique of Sec.~\ref{sec:lkh} & 1.60	\\ \hline
Monte Carlo likelihood technique & 1.37		\\ \hline
Feldman \& Cousins~\cite{feldman}	& $\sim 2.5$	\\ \hline
\end{tabular}
\end{table}
%\end{center}
The value given by the procedure~\cite{feldman} of Feldman and
Cousins is a rough estimate only, since the combination of input
parameters $\{n=3,b=5.5\}$ is not shown in
their tables.

To implement the maximum likelihood approach of Section~\ref{sec:lkh},
the signal Monte Carlo two-dimensional distribution on the energy-vs-mass plane
was fitted to a bivariate Gaussian plus a non-Gaussian tail produced
by initial and final state radiation and other effects:
\begin{eqnarray}
\ss_i (m,E) & = & \frac{1}{A+B} \left\{
\frac{A}{2\pi\sigma_m\sigma_E\sqrt{1-\rho^2}} \right. \times \nonumber \\
& & \left. \times exp
\left\{
-\frac{1}{2(1-\rho^2)}
\left[ 
\left( \frac{m-m_0}{\sigma_m} \right)^2
- 2\rho \left(\frac{m-m_0}{\sigma_m}\right)\left(\frac{E-E_0}{\sigma_E}\right)
+ \left( \frac{E-E_0}{\sigma_E} \right)^2
\right]
\right\} \right. + \nonumber \\
& & \left. + B\epsilon(m,E) \right\}\ ;\\
& & \nonumber \\
\epsilon(m,E) & = & \left\{
\begin{array}{cl}
\frac{1}{\sqrt{2\pi}\sigma_m}
exp\left[ -\frac{1}{2} \left( \frac{m-m_0}{\sigma_m} \right)^2 \right]
\frac{1}{\sigma_E\Gamma(\alpha)\beta^\alpha}
\left( \frac{E_0-E}{\sigma_E} \right)^{\alpha-1}
exp\left[ -\frac{E_0-E}{\beta\sigma_E} \right] & \mbox{if $E<E_0$} \\
0 & \mbox{otherwise}
\end{array}
\right. \nonumber
\end{eqnarray}
where $A$, $B$, $\sigma_m$, $\sigma_E$, $\rho$, $m_0(\approx m_\tau)$,
$E_0(\approx E_{beam})$, $\alpha$ and $\beta$ are the fit parameters.
The background spatial pdf $\bb_i$ was obtained by fitting
data events observed in the vicinity of the signal region to a linear
function: 
\begin{equation}
\bb_i (m,E) = \frac{1}{m_2-m_1} 
\frac{1}{ (a_0-a_1E')(E_2-E_1) + 0.5a_1(E_2^2-E_1^2) }
\left[ a_0 + a_1(E-E') \right]\ ,
\end{equation}
where $a_0$, $a_1$ and $E'$ are the fit parameters, and $(m_1,m_2)$
and $(E_1,E_2)$ are the limits defining the fit region.
The three events observed in the signal region are located far from
the peak of the signal Monte Carlo distribution. Thus, the maximum
likelihood fit treats these events as background, and the extracted
signal rate is consistent with zero. This explains why the likelihood
integration technique gives the value of 2.30.

\section{Conclusion}

There is no such thing as the ``best'' procedure for upper limit
estimation. An experimentalist is free to choose any procedure she/he
likes, based on her/his belief and experience. The only requirement is
that the chosen procedure must have a strict mathematical
foundation. The Bayesian method with a flat prior and the
likelihood integration technique of Section~\ref{sec:lkh} seem to have
been the two most popular choices in the past few years. Typically, these two
approaches give the most conservative values of upper limits. The
purpose of this note is to show that there are other approaches,
equally justified by mathematical formalism, which produce less
conservative estimates.

\section{Acknowledgements}

The author thanks Prof.~W.T.~Ford, Prof.~R.~Stroynowski,
Prof.~H.L.~Gray and Dr.~I.~Volobouev for useful
comments and suggestions.

%\end{spacing}

\end{document}